# Demonstration of Beyond Terabit/s/λ Nonlinearity-free Transmission over the Hollow-core Fibre

Yang Hong[(1)], Sylvain Almonacil[(1)], Haik Mardoyan[(1)], Carina Castineiras Carrero[(1)], Sergio Osuna[(2)], Javier R. Gomez[(3)], David R. Knight[(4)], Jeremie Renaudier[(1)]

[(1)] Optical Transmission Dept, Nokia Bell Labs, Massy, France, yang.hong@nokia.com
[(2)] Nokia, Networks Infrastructure Division, Madrid, Spain
[(3)] Lyntia, Madrid, Spain
[(4)] OFS Laboratories, Somerset, USA

**Abstract** *We demonstrate nonlinearity-free transmission of Terabit/s/λ PCS-64QAM signals through an HCF-based optical recirculating loop, which yields ~17.4% higher capacity than SMF-based loop under 23-dBm launch power (~13.5 dBm/channel) after 25 loops. Both lab experiment and field trial show HCF exhibits ~1.6-μs/km lower latency than SMF. ©2024 The Author(s)*

**Introduction**

Driven by the emerging data-intensive and latency-sensitive applications such as cloud/edge computing and ultrahigh-definition video streaming, current optical fibre communication empowered by the conventional solid-core single-mode fibre (SMF) is facing a persistent demand for higher capacity and lower latency [1,2]. Over the past years, substantial capacity improvements have been achieved using various solutions, including advanced digital signal processing (DSP) [3,4], ultra-wideband transmission [5], and space division multiplexing [6]. However, the achievable capacity of the solid-core SMF is inevitably limited by its inherent fibre nonlinearity, which is referred to as the nonlinear Shannon limit [7]. On the other hand, despite that transmission fibre represents the major latency contributor in the overall system (even for short-reach scenarios) [8], latency reduction of SMF to date has been minimal (only 0.3% reduction have been reported [9]). This is because light speed in SMF is ultimately limited by the refractive index of the silica-glass core.

Hollow-core fibres (HCFs) that rely on a fundamentally different light guiding mechanism offer a viable solution to simultaneously overcome the nonlinearity and latency limitations of SMF [10-18]. The unique air-guided light propagation property enables HCFs to be the ultimate low-latency transmission medium [11-13] with the lowest reported propagation loss [14] amongst any types of optical fibres. In the meantime, HCFs also exhibit a much lower fibre nonlinearity (around three to four orders of magnitude) than the solid-core SMF [15], thereby exhibiting the potential to realize nonlinearity-free transmission (in other words, break the fundamental nonlinear Shannon limit experienced in SMFs). In the literature, several transmission experiments using high launch powers to reflect the ultra-low nonlinearity of HCFs have been reported [16-18]. In [16], a 11.5-km HCF was adopted in an optical recirculating loop with a launch power of 20 dBm (the launch power of each channel was ~3.9 dBm), supporting the transmission of 32-GBaud dual-polarization (DP) signals. To directly showcase the low nonlinearity of HCF relative to SMF, both types of fibres were tested by three-channel 400G signals over a link of ~10 km. It was shown that the HCF could enable nonlinearity-penalty-free transmission with a total launch power up to 20.3 dBm, whereas SMF only tolerated up to ~15 dBm [17]. In [18], penalty-free transmission of 95-GBaud DP probabilistic constellation shaping (PCS) 64QAM signals (800Gb/s/λ) was achieved with a total launch power up to 28 dBm (which corresponded to around 12 dBm/channel). However, only a straight-line link with a short length (~200 m) was considered.

In this paper, with an optical recirculating loop, we show that HCF supports nonlinearity-free transmission of Terabit/s/λ 130-GBaud PCS-64QAM signals with a total launch power up to 23 dBm (~13.5 dBm/channel). Furthermore, it is validated through both lab experiment and field trial that HCF offers ~1.6-μs/km latency reduction (~32%) relative to SMF. To the best of our knowledge, our work constitutes the first Terabit/s/λ transmission (and the highest baud rate ever reported) over the HCF, highlighting the potential of incorporating HCFs with the state-of-the-art high baud-rate coherent transceivers for future low-latency and ultrahigh-capacity optical communications.

**Experimental Setup**

Fig. 1(a) shows the experimental setup of the optical recirculating loop. At the transmitter, the 130-GBaud channel under test (CUT) centred at 1559.39 nm was generated by modulating a DP coherent driver modulator (DP-CDM) with a 256-GS/s arbitrary waveform generator (AWG, 10-dB bandwidth is ~70 GHz). The output of a C-band amplified spontaneous emission (ASE) source was first amplified by an erbium-doped optical fibre amplifier (EDFA), before been spectrally shaped by a WaveShaper (WS) to generate eight 130-GBaud neighbouring comb channels with a channel spacing of 150 GHz. The WDM comb and the amplified CUT were combined together and then further amplified by another EDFA, before been fed into the optical recirculating loop. Fig. 1(b) shows the spectra of the WDM channels at the transmitter. The spectral range was chosen to match the low-loss window of the adopted HCF, as will be presented later.

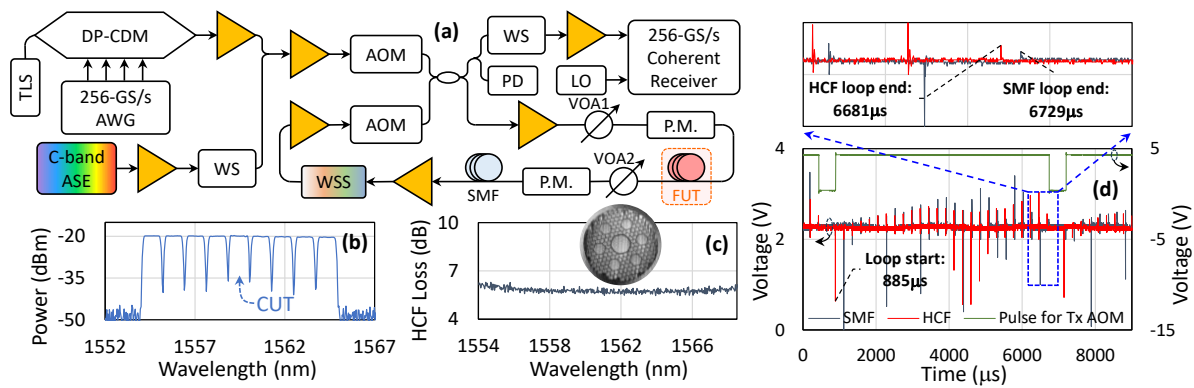

**Fig. 1:** (a) Experimental setup, (b) optical spectra of the CUT and comb channels at the transmitter, (c) the insertion loss of the 1.085-km length of HCF (including SMF-HCF interconnection loss), and (d) comparison of the power monitoring traces of the HCF-/SMF-based loop after 25 loops. Inset of (c): cross-sectional scanning electron microscope image of the used HCF.

The loop was composed by two acousto-optic modulators (AOMs) and a 3-dB optical coupler. In the loop, a high-power EDFA was used to boost the launch power into the fibre under test (FUT), which was controlled by a variable optical attenuator (VOA1) and monitored by a power meter (P.M.). We considered both HCF and SMF as the FUT (around 1.1 km) in the loop. The total insertion loss of the SMF-connectorized HCF was measured to be around 5.7 dB at the wavelengths of interest, as shown in Fig. 1(c). The HCF adopted the photonic bandgap design with six surrounding shunt cores to suppress the higher-order modes and multipath interference from modal crosstalk [13]. After the FUT, another VOA (VOA2) was used to fix the optical power into the subsequent buffering SMF (~45.6 km) to be 10.2 dBm, at which the impact of the buffering SMF's nonlinearity could be considered as negligible. Finally, a 50-GHz-grid wavelength-selective switch (WSS) and a pair of EDFAs were used to equalize the WDM channels and balance the loss and gain inside the loop which was monitored by a photodetector (PD) at the receiver side. The output of the loop was filtered by a WS to select the CUT. The resulting optical signal was then amplified before fed into a standard coherent receiver front-end, the output of which was captured by a 256-GS/s real-time oscilloscope (3-dB bandwidth is ~100 GHz) for offline DSP.

In this work, the PCS-64QAM with an entropy of 5.7 bit/symbol was adopted as the modulation format. To combat the limited bandwidth of the transmitter, digital pre-emphasis was applied. At the receiver side the captured signal was processed by standard DSP as in [19]. We note that no digital compensation for fibre nonlinearity was performed in the offline DSP. Finally, the de-modulated signals were used to evaluate the signal-to-noise ratio (SNR) and the achievable information rate (AIR).

**Experimental Results**

We first evaluated the latency performance of the transmission by adopting either HCF or SMF as the FUT (~1.1 km) in the loop (together with the around 45.6-km buffering SMF), which are referred to as 'HCF loop' and 'SMF loop' later, respectively. Fig. 1(d) shows the power monitoring traces in the two cases after 25 loops (the number of recirculation loops was controlled by the control pulse for the AOM as illustrated in Fig. 1(d)). The spikes indicate the end of each loop, using which the latency difference between HCF and SMF can be directly assessed. It is seen that the total propagation time of the HCF/SMF loops are 5,796 μs and 5,844 μs, and the corresponding fibre latencies of the HCF and SMF are then calculated as ~3.3 μs/km and ~5 μs/km, respectively. This validates that HCF exhibits a significantly lower latency when compared to SMF.

We further varied the launch power into the FUT (i.e., HCF/SMF) in the loop to investigate the impact of its nonlinearity on the transmission performance. As aforementioned, VOA2 was adjusted to fix the power to the subsequent buffering SMF to 10.2 dBm in all cases, regardless of the varying launch power into the FUT. As such, if the FUT's nonlinearity is sufficiently low, under the same number of loops, similar transmission performance should be expected at different launch powers. As shown in Fig. 2(a), for the 'HCF loop', the SNRs remain comparable at all tested launch powers (up to 23 dBm), as expected. Furthermore, this is achieved at different numbers of loops, which clearly indicates that no SNR penalty was induced by the ~1.1-km HCF, thanks to its inherent ultra-low nonlinearity. It is worth noting that the maximum total launch power (i.e., 23 dBm, which corresponds to ~13.5 dBm/channel) was limited by the available high-power EDFA when performing the experiments. We anticipate that the HCF should be able to support nonlinearity-free transmission with even higher launch powers.

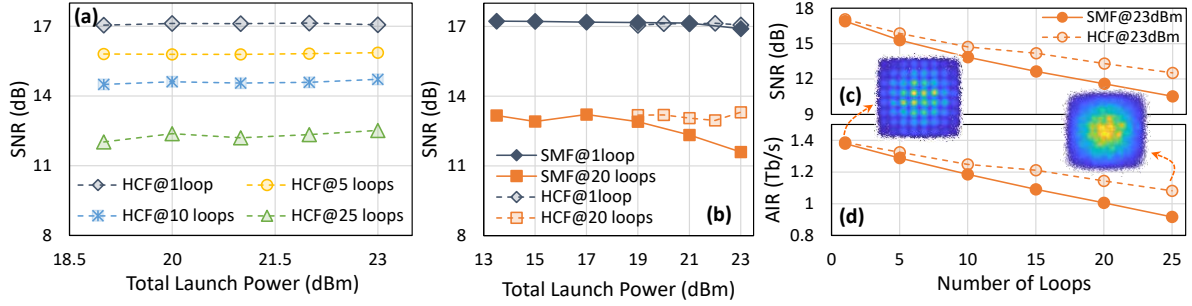

**Fig. 2:** (a) SNR versus launch power in the 'HCF loop' case, (b) comparison of the AIR versus launch power between the 'HCF loop' and 'SMF loop' cases, (c) SNR versus number of loops, and (d) AIR versus number of loops. Note that at the same number of loops, constant OSNR was maintained by adjusting VOA2 despite the increase in launch power into the FUT. Insets: the constellation diagrams of the PCS-64QAM after 1 loop and 25 loops, respectively, in the 'HCF loop' case.

In comparison, as shown in Fig. 2(b), when the FUT was changed to ~1.1-km SMF, similar SNRs can also be achieved at all tested launch powers in the scenario of 1-loop transmission. This indicates that the FUT's nonlinearity (in this case, SMF) is negligible after only 1 loop, since its length is just ~1.1 km. However, when the number of loops is increased to 20, similar SNRs of around 13.2 dB can only be maintained by keeping the total launch power below 19 dBm. Further increasing the launch power results in a significant decrease in SNR due to the accumulated FUT's nonlinearity after 20 loops.

We then kept the total launch power at 23 dBm (i.e., ~13.5 dBm/channel) and investigated the impact of FUT's nonlinearity after different numbers of loops. As shown in Fig. 2(c), a relatively minor SNR penalty is observed in the 'SMF loop' case if the number of travelled loops is small. However, the SNR penalty relative to the 'HCF loop' case tends to be significant when further increasing the number of loops. This results from the accumulated nonlinearity arising from the ~1.1-km SMF (i.e., FUT). As a result, a SNR improvement of ~2 dB is achieved after 25 loops. Accordingly, the AIR performance follows the same trend as that of the SNR results. As depicted in Fig. 2(d), after 25 loops, the AIR can be improved from ~0.92 Tb/s to ~1.08 Tb/s by employing the HCF instead of SMF as the FUT. This corresponds to ~17.4% capacity enhancement

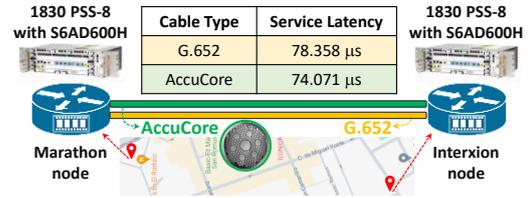

**Fig. 3:** Configuration of the field trial.

thanks to the ultra-low nonlinearity of the HCF.

Finally, a field trial was also conducted in Lyntia's backbone network in Madrid, Spain, from Marathon node to Interxion. The configuration of the trial is illustrated in Fig. 3. Nokia 1830 PSS-8 platforms with S6AD600H transponders were adopted to realize 90.22-GBaud 16QAM transmission (600G) at 1562.79 nm, enabling 100G real-time Ethernet services. A HCF (AccuCore) and a SMF (G.652, with a matched loss with the HCF using a VOA) were packed in the same ~1.4-km *cable* and deployed in the field. While similar BER performance was achieved by the HCF and SMF links, the HCF link exhibited a round-trip latency reduction of 4.287 μs for the 100G services. Since both the HCF and SMF are within the same cable, their physical length is known to be almost identical. Therefore, the resulting ~1.6-μs/km latency reduction is independent of length related errors. The field trial clearly indicates the practical viability of using HCFs for latency-sensitive connectivity, e.g., data center and high-frequency trading applications.

## Conclusions

We demonstrate that the use of HCF allows for nonlinearity-free transmission of Terabit/s/λ 130-GBaud PCS-64QAM signals at launch powers up to 23 dBm (which corresponds to ~13.5 dBm/channel). After 25 recirculation loops, this leads to ~2-dB higher SNR and thus ~17.4% capacity improvement relative to the case of using SMF. Furthermore, it is verified through both lab experiment and field trial that HCF exhibits ~1.6-μs/km (~32%) lower latency than SMF.

## References

[1] P.J. Winzer, D.T. Neilson, and A.R. Chraplyvy, "Fiber-optic Transmission and Networking: The Previous 20 and The Next 20 Years," *Optics Express*, vol. 26, no. 18, pp. 24190-24239, 2018, DOI: 10.1364/OE.26.024190.

[2] P. Poggiolini, and F. Poletti, "Opportunities and Challenges for Long-Distance Transmission in Hollow-Core Fibres," *Journal of Lightwave Technology*, vol. 40, no. 6, pp. 1605-1616, 2022, DOI: 10.1109/JLT.2021.3140114.

[3] T. Fehenberger, A. Alvarado, G. Böcherer, N. Hanik, "On Probabilistic Shaping of Quadrature Amplitude Modulation for the Nonlinear Fiber Channel," *Journal of Lightwave Technology*, vol. 34, no. 21, pp. 5063-5073, 2016, DOI: 10.1109/JLT.2016.2594271.

[4] F. Musumeci, C. Rottondi, A. Nag, I. Macaluso, D. Zibar, M. Ruffini, M. Tornatore, "An Overview on Application of Machine Learning Techniques in Optical Networks," *IEEE Communications Surveys & Tutorials*, vol. 21, no. 2, pp. 1383-1408, 2019, DOI: 10.1109/COMST.2018.2880039.

[5] B.J. Puttnam, R.S. Luís, G. Rademacher, M. Mendez-Astudillio, Y. Awaji, and H. Furukawa, "S-, C- and L-band Transmission over a 157 nm Bandwidth using Doped Fiber and Distributed Raman Amplification," *Optics Express*, vol. 30, no. 6, pp. 10011-10018, 2022, DOI: 10.1364/OE.448837.

[6] W. Klaus, P.J. Winzer, and K. Nakajima, "The Role of Parallelism in the Evolution of Optical Fiber Communication Systems," *Proceedings of the IEEE*, vol. 110, no. 11, pp. 1619 -1654, 2022, DOI: 10.1109/JPROC.2022.3207920.

[7] A.D. Ellis, J. Zhao, and D. Cotter, "Approaching the Non-Linear Shannon Limit," *Journal of Lightwave Technology*, vol. 28, no. 4, pp. 423-433, 2010, DOI: 10.1109/JLT.2009.2030693.

[8] Y. Hong, K.R.H. Bottrill, T.D. Bradley, H. Sakr, G.T. Jasion, K. Harrington, F. Poletti, P. Petropoulos, and D.J. Richardson, "Low-Latency WDM Intensity-Modulation and Direct-Detection Transmission Over >100 km Distances in a Hollow Core Fiber," *Laser & Photonics Reviews*, vol. 15, paper 2100102, 2021, DOI: 10.1002/lpor.202100102.

[9] Y. Sagae, T. Matsui, K. Tsujikawa, and K. Nakajima, "Solid-Type Low-Latency Optical Fiber With Large Effective Area," *Journal of Lightwave Technology*, vol. 37, no. 19, pp. 5028-5033, 2019, DOI: 10.1109/JLT.2019.2927168.

[10] D.J. Richardson, "Recent Advances in Hollow-Core Optical Fibers," in *Proceedings of 27th OptoElectronics and Communications Conference* (OECC), Toyama, Japan, paper MC3-1, 2022, DOI: 10.23919/OECC/PSC53152.2022.9849996.

[11] Y. Hong, T.D. Bradley, N. Taengnoi, K.R.H. Bottrill, J.R. Hayes, G.T. Jasion, F. Poletti, P. Petropoulos, and D.J. Richardson, "Hollow-Core NANF for High-Speed Short-Reach Transmission in the S+C+L-Bands," *Journal of Lightwave Technology*, vol. 39, no. 19, pp. 6167-6174, 2021, DOI: 10.1109/JLT.2021.3097278.

[12] A.C. Meseguer; J.T. de Araujo. and J.-C. Antona, "Multi-Core vs Hollow-Core Fibers: Technical Study of Their Viability in SDM Power-Constrained Submarine Systems," *Journal of Lightwave Technology*, vol. 41, no. 12, pp. 4002-4009, 2023, DOI: 10.1109/JLT.2023.3278714.

[13] B. Zhu, B.J. Mangan, T. Kremp, G.S. Puc, V. Mikhailov, K. Dube, Y. Dulashko, M. Cortes, Y. Liang, K. Marceau, B. Violette, D. Cartsounis, R. Lago, B. Savran, D. Inniss, and D.J. DiGiovanni, "First Demonstration of Hollow-Core-Fiber Cable for Low Latency Data Transmission," in *Proceedings of Optical Fiber Communication Conference (OFC)*, San Diego, United States, paper Th4B.3, 2020, DOI: 10.1364/OFC.2020.Th4B.3.

[14] Y. Chen, M.N. Petrovich, E.N. Fokoua, A.I. Adamu, M.R.A. Hassan, H. Sakr, R. Slavík, S.B. Gorajoobi, M. Alonso, R.F. Ando, A. Papadimopoulos, T. Varghese, D. Wu, M.F. Ando, K. Wisniowski, S.R. Sandoghchi, G.T. Jasion, D.J. Richardson, F. Poletti, "Hollow Core DNANF Optical Fiber with <0.11 dB/km Loss," in *Proceedings of Optical Fiber Communication Conference (OFC)*, San Diego, United States, paper Th4A.8, 2024.

[15] D. Ge, S. Gao, M. Zuo, Y. Gao, Y. Wang, B. Yan, B. Ye, D. Zhang, W. Ding, H. Li, and Z. Chen, "Estimation of Kerr Nonlinearity in an Anti-resonant Hollow-Core Fiber by High-order QAM Transmission," in *Proceedings of Optical Fiber Communication Conference (OFC)*, San Diego, United States, paper W4D.6, 2023, DOI: 10.1364/OFC.2023.W4D.6.

[16] A Nespola; S R Sandoghchi; L Hooper; M Alonso; T D Bradley; H Sakr; G T Jasion; E Numkam Fokoua; S Straullu, G Bosco, A Carena, Y Jiang, A M Rosa Brusin, Y Chen, J R Hayes, F Forghieri, D J Richardson, F Poletti, P Poggiolini, "Ultra-Long-Haul WDM PM-16QAM Transmission in a Reduced Inter-Modal Interference NANF," in *Proceedings of IEEE Photonics Conference (IPC)*, Orlando, USA, paper WB1.3, 2023, DOI: 10.1109/IPC57732.2023.10360760.

[17] A. Iqbal, P. Wright, N. Parkin, M. Fake, M. Alonso, S.R. Sandoghchi, and A. Lord, "First Demonstration of 400ZR DWDM Transmission through Field Deployable Hollow-Core-Fibre Cable," in *Proceedings of Optical Fiber Communication Conference (OFC)*, San Francisco, United States, paper F4C.2, 2021, DOI: 10.1364/OFC.2021.F4C.2.

[18] D. Ge, S. Gao, M. Zuo, Y. Gao, Y. Wang, D. Zhang, W. Ding, H. Li, X. Duan, and Z. Chen, "Nonlinear-penalty-free Real-time 40×800Gb/s DP-64QAM-PCS Transmission with Launch Power of 28 dBm over a Conjoined-tube Hollow-core Fiber," in *Proceedings of Optical Fiber Communication Conference (OFC)*, San Diego, United States, paper W4H.7, 2023, DOI: 10.1364/OFC.2023.W4H.7.

[19] A. Ghazisaeidi, I.F.J. Ruiz, R. Rios-Müller, L. Schmalen, P. Tran, P. Brindel, A.C. Meseguer, Q. Hu, F. Buchali, G. Charlet, and J. Renaudier, "Advanced C+L-Band Transoceanic Transmission Systems Based on Probabilistically Shaped PDM-64QAM," *Journal of Lightwave Technology*, vol. 35, no. 7, pp. 1291-1299, 2017, DOI: 10.1109/JLT.2017.2657329.